\documentclass[a4]{epl2}
\usepackage{graphicx}
\usepackage{epsfig}
\usepackage{color}
\usepackage{txfonts}

\title{A novel model for the fractional quantum Hall effect}
\shorttitle{A novel model for the fractional quantum Hall effect}
\author{A. I. Arbab\inst{}\footnote{aiarbab@uofk.edu}}

\institute{
  \inst{} Department of Physics,
Faculty of Science, University of Khartoum, P.O. Box 321, Khartoum
11115, Sudan
}

\abstract{A novel model of complex quantum harmonic oscillator is found to account for the observed Fractional quantum Hall effect (FQHE). The sequences of the observed FQHE conductivity and charge are explained. The two sequences are found to express a quantity and its complex conjugated partner. The oscillator is found to have two degenerates states, $\psi_n$,  with  angular momenta $\pm \,n\,\hbar$\,, where $h = 2\pi \hbar $ is the Planck's constant, and $n$ is the principal quantum number of the oscillator. The filling factor, $i$, that Klitzing has found for the integer quantum Hall effect (IQHE) is $i=n+1$. Analytical expressions for longitudinal resistance and Hall's voltage are obtained.  The width of the plateau between two states is found to be $\Delta B=\frac{1}{n(n+1)}\,\frac{n_sh}{e}\,,$ where $n_s$ is the electron number density.}

\begin{document}

\maketitle
\section{Introduction}
A classical  Hall effect was introduced by Edwin Hall in 1879 to account for the accumulation of charges when a transverse magnetic field is applied to a conductor \cite{1}. As a result a transverse (Hall) resistance and voltage are then developed.   This effect was later observed for semiconductors.  The resistivity (conductivity) is found to vary with the applied magnetic field. A quantum Hall effect is however present  at high magnetic field. In such situations a series of plateaus form in the Hall conductivity that are  multiples of $e^2/h$, where $e$ and $h$ are the electronic charge and Planck's constant \cite{kilit}.
A fractional quantum Hall effect (FQHE)  is observed when the Hall conductivity of 2D electrons gas shows a  quantized plateaus at fractional values multiple of $e^2/h$ \cite{tsui}. Each particular value of the magnetic field corresponds to a filling factor (the ratio of electrons to magnetic flux quanta), $\nu=p/q$, where $p$ is an integer, and $q$ is an odd integer. This effect occurs at high magnetic field and low temperature. This behavior is found to be a manifestation of  a collective state where electrons bind magnetic flux lines to make new quasiparticles (excitations) \cite{laugh}. These excitations have a fractional elementary charge that is supposed to obey some fractional statistics \cite{jain}. The FQHE was experimentally discovered by Daniel Tsui and Horst St\"{o}rmer, in experiments performed on gallium arsenide heterostructures developed by Arthur Gossard Tsui, St\"{o}rmer, and Laughlin \cite{tsui}.
Fractionally charged quasiparticles are neither bosons nor fermions and exhibit anyonic statistics \cite{leinas}. Laughlin proposed  trial wave functions for the ground state at fraction of $1/q$, as well as, its quasiparticle and quasihole excitations. These excitations have fractional charge of magnitude $e^*=e/ q$. The most probable theory to explain the fractional quantum Hall effect is the composite fermion model introduced by  Jain \cite{jain}.

Alkane \cite{hald} and later  Haltering \cite{halp} proposed a hierarchy scheme to explain the observed filling fractions not occurring at the Laughlin states\cite{laugh}.  They have shown that starting with the Laughlin states, new states at different fillings can be formed by condensing quasiparticles into their own Laughlin states. The new states and their fillings are constrained by the fractional statistics of the quasiparticles \cite{jain}. In this manner, all hierarchy of states covering all the odd-denominator filling fractions is  produced. Furthermore,  a composite fermions model was proposed by Jain \cite{jain}, and extended by Haltering \emph{et al.} \cite{jain,read}. This is based on the idea that composite fermions are formed from integer-charged quasiparticles as a result of the repulsive interactions, where two  vortices are captured by each electron. Composite fermions have been observed, and the theory has been experimentally verified \cite{Du}.

Thus, the  composite fermion model provides a complementary description of the Laughlin and hierarchy states \cite{halp}. It gives trial wave functions which, though not identical to those produced from the hierarchy picture, are in the same universality class, as shown by Read \cite{read}.

We provide in this paper a new formulation to explain the fractional quantum Hall effect. This model can be compared with the composite model of Jain \cite{jain}. To this aim, we employ a complex harmonic oscillator which is equivalent to a two dimensional harmonic oscillator \cite{arbab}. In this formulation each oscillator (particle) is represented by four virtual  states in the complex plane. These are performed by continuous reflection and rotation of a given point in the complex plane. Thus, a given particle is represented by a quartet of virtual particles. The states describing the oscillator are degenerate but with different angular momenta, \emph{viz.,} $\pm n\hbar$. We call these two states the conjugate states. The two observed series in the fractional quantized Hall conductivity arises from these two degenerate states.
\section{Complex quantum Harmonic oscillator}
The Hamiltonian and  the angular momentum of the complex quantum harmonic oscillator are hermitian (real), i.e., $\bar{H}_z=H_z$ and $\bar{L}_z=L_z$.
The Hamiltonian of the 2-dimensional system is given by \cite{arbab}
\begin{equation}\label{1}
H_{z\bar{z}}=(\bar{a}_za_z+1)\,\hbar\,\omega+\omega\,L_z\,,
\end{equation}
where the number operator is
\begin{equation}\label{1}
N_z=\bar{a}_za_z\,,
\end{equation}
so that eq.(1) becomes
\begin{equation}\label{1}
H_{z\bar{z}}=(N_z+1)\,\hbar\,\omega+\omega\,L_z\,.
\end{equation}
The solution of Schrodinger equation yields two complex wavefunctions. These are defined by \cite{arbab}
\begin{equation}
 \psi_{2 n}=C_{2 n}z^{\,n}e^{-\alpha z\bar{z}}\,,\,\,\,  \psi_{1 n}=C_{1 n}\bar{z}^{\,n}e^{-\alpha z\bar{z}}\,,
\end{equation}
where $C_{1n}$, $C_{2n}$ and $\alpha=\frac{m\omega}{2\hbar}$ are constants.
These two wavefunctions can be normalized as \cite{arbab}
$$
\int|\psi_{2n}(z, \bar{z})|^2dx\,dy=\int|\psi_{1 n}(z, \bar{z})|^2dx\,dy=1\,\,,$$
 \begin{equation}
C_{1n}=C_{2 n}=\sqrt{\frac{(2\alpha)^{n+1}}{\!\!\!2\pi\,n!}}\,,
\end{equation}
where  $z=re^{i\theta}$, $dxdy=rdrd\theta$.

The expectation values of the  momentum in the state $\psi_{2 n}$ are  given by
\begin{equation}
 <p_z>=-\frac{(2n-1)}{2\pi}\,\hbar\,\,\,\sqrt{\frac{m\omega}{\hbar}}\,\,\,\frac{\Gamma(n+\frac{1}{2})}{\Gamma(n+1)}\,,
 \end{equation}
 and
 \begin{equation}
 <p_{\bar{z}}>=-\frac{(2n+1)}{2\pi}\,\hbar\,\,\,\sqrt{\frac{m\omega}{\hbar}}\,\,\,\frac{\Gamma(n+\frac{1}{2})}{\Gamma(n+1)}\,,
 \end{equation}
 whereas the expectation values of the  momentum in the states  $\psi_{1 n}$ are  given by
 \begin{equation}
 <p_z>=\frac{(2n+1)}{2\pi}\,\hbar\,\,\,\sqrt{\frac{m\omega}{\hbar}}\,\,\,\frac{\Gamma(n+\frac{1}{2})}{\Gamma(n+1)}\,,
 \end{equation}
 and
 \begin{equation}
 <p_{\bar{z}}>=\frac{(2n-1)}{2\pi}\,\hbar\,\,\,\sqrt{\frac{m\omega}{\hbar}}\,\,\,\frac{\Gamma(n+\frac{1}{2})}{\Gamma(n+1)}\,,
 \end{equation}
 respectively.
 The eigen values of the angular momentum of the two states are $n\hbar$ and $-n\hbar$, i.e., \cite{arbab}
 \begin{equation}\label{5}
L_z\psi_{2 n}=n\hbar\,\psi_{2 n}\,,\qquad     L_z\psi_{1 n}=-n\hbar\,\psi_{1 n}\,.
 \end{equation}
We observe here that the complex harmonic oscillator is a quantum system where its energy as well as its angular momentum are quantized. We may associate the states,  $L_z=n\hbar$ and $L_z =-n\hbar$, with a particle rotating clockwise and counterclockwise, respectively. Furthermore, notice from eqs.(6) - (9) that the $<p_z>$ of $\psi_{1 n}$ is equal to $- <p_{\bar{z}}>$ of $\psi_{2 n}$\,, and $<p_{\bar{z}}>$ of $\psi_{1 n}$ is equal to $- <p_{z}>$ of $\psi_{2 n}$\,.

The two states $\psi_{1 n}$ and $\psi_{2 n}$ are degenerate, since
 \begin{equation}
H_{z\bar{z}}\,\psi_{2 n}=(n+1)\hbar\,\omega\,\psi_{2 n}\,,\qquad     H_{z\bar{z}}\,\psi_{1 n}=(n+1)\hbar\,\omega\,\psi_{1 n}\,.
 \end{equation}
 The maximum of the above squared wavefunction for any state ($n$) occurs at (where $z=re^{i\theta}$)
  \begin{equation}\label{5}
r_{0n}=\sqrt{\frac{\hbar}{m\omega}}\, \sqrt{n}\,\,, \qquad n=0, 1, 2, \cdots
 \end{equation}
 For landau level, i.e., the degenerate levels  resulting from  the energy spectrum of a charged particle moving in a constant magnetic field, the  magnetic length is defined as, $l_B=\sqrt{\frac{\hbar}{m\omega_c}}=\sqrt{\frac{\hbar}{eB}}$\,, where $\omega_c=\frac{eB}{m}$ is the cyclotron frequency of the electron \cite{landau}. For this case, $r_{0 n}=\sqrt{n}\,\,\,l_B$. Moreover, the  angular momentum of the classical cyclotron motion is, $L_c=mvr=eBr^2$. For the above maximum state, i.e., eq.(12), one has $L_c=n\hbar$. This agrees with the quantum value.

\section{Electric and magnetic properties of the oscillators}
Let us now define the xy  currents (per  unit length) in the state $\psi_{2 n}$ as \cite{arbab}
$$
{\cal J}_x=\frac{q\hbar}{2mi}\left(\bar{\psi}_{2 n}\frac{\partial\psi_{2 n}}{\partial x}-\psi_{2 n}\frac{\partial\bar{\psi}_{2 n}}{\partial x}\right)\,,$$ \begin{equation}{\cal J}_y=\frac{q\hbar}{2mi}\left(\bar{\psi}_{2 n}\frac{\partial\psi_{2 n}}{\partial y}-\psi_{2 n}\frac{\partial\bar{\psi}_{2 n}}{\partial y}\right)\,\,,
 \end{equation}
 which yield
$$
{\cal J}_x=-\frac{nq\hbar}{2im}\,(\bar{z}-z)\,C_1^2\,|z|^{2n-2}\,\,e^{-2\alpha|z|^2}\,,$$
 \begin{equation}
 {\cal J}_y=\frac{nq\hbar}{2m}\,(z+\bar{z})\,C_1^2\,|z|^{2n-2}\,\,e^{-2\alpha|z|^2}\,,
 \end{equation}
  and in the state $\psi_{1 n}$ as
$$
{\cal J}_x=\frac{q\hbar}{2mi}\left(\bar{\psi}_{1 n}\frac{\partial\psi_{1 n}}{\partial x}-\psi_{1 n}\frac{\partial\bar{\psi}_{1 n}}{\partial x}\right)\,,$$
  \begin{equation}
 {\cal J}_y=\frac{q\hbar}{2mi}\left(\bar{\psi}_{1 n}\frac{\partial\psi_{1 n}}{\partial y}-\psi_{1 n}\frac{\partial\bar{\psi}_{1 n}}{\partial y}\right)\,\,,
 \end{equation}
 which yield
$$
{\cal J}_x=-\frac{nq\hbar}{2im}\,(\bar{z}-z)\,C_2^2\,|z|^{2n-2}\,\,e^{-2\alpha|z|^2}\,,$$
 \begin{equation}
 {\cal J}_y=\frac{nq\hbar}{2m}\,(z+\bar{z})\,C_2^2\,|z|^{2n-2}\,\,e^{-2\alpha|z|^2}\,.
 \end{equation}
Thus, if the above particle (oscillator) is an electron, then $4q=e$ \cite{arbab}.
 In terms of real coordinates, one has the corresponding currents in the states $\psi_{2 n}$ and $\psi_{1 n}$\,, respectively, as
 $$
{\cal J}_x=-\frac{ne\hbar}{4m}\,C_2^2\,r^{2n-1}\,\,e^{-2\alpha \,r^2}\,\sin\theta\,,  $$
  \begin{equation}
 {\cal J}_y=\frac{ne\hbar}{4m}\,C_2^2\,r^{2n-1}\,\,e^{-2\alpha \,r^2}\,\cos\theta\,,
 \end{equation}
 and
$${\cal J}_x=\frac{ne\hbar}{4m}\,C_2^2\,r^{2n-1}\,\,e^{-2\alpha \,r^2}\,\sin\theta\,, $$
  \begin{equation}
 {\cal J}_y=\frac{ne\hbar}{4m}\,C_2^2\,r^{2n-1}\,\,e^{-2\alpha \,r^2}\,\cos\theta\,.
 \end{equation}
 Thus, integrating eq.(17) yields the current moments  along the x- and y- directions as
 \begin{equation}\label{5}
{\cal I}_x=-\frac{ne\hbar}{2\pi\, m}\,\,\sqrt{\frac{m\,\omega}{ \hbar}}\,\,\,\frac{\Gamma(n+\frac{1}{2})}{\Gamma(n+1) }\,\,, {\cal I}_y=0\,, n=0, 1, 2, \cdots\,.
 \end{equation}
 Hence, the current in the state $\psi_{2 n}$ is
 \begin{equation}\label{5}
{\cal I}_{z, 2 n}={\cal I}_x+i\,{\cal I}_y=-\frac{ne\hbar}{2\pi\, m}\,\,\sqrt{\frac{m\,\omega}{ \hbar}}\,\,\,\frac{\Gamma(n+\frac{1}{2})}{\Gamma(n+1) }\,.
 \end{equation}
Integrating eq.(18) yields the current moments  along the x- and y- directions as
 \begin{equation}\label{5}
{\cal I}_x=\frac{ne\hbar}{2\pi\, m}\,\,\sqrt{\frac{m\,\omega}{ \hbar}}\,\,\,\frac{\Gamma(n+\frac{1}{2})}{\Gamma(n+1) }\,\,,  {\cal I}_y=0\,,\, n=0, 1, 2, \cdots\,.
 \end{equation}
 Hence, the current in the state $\psi_{1 n}$ is
 \begin{equation}\label{5}
{\cal I}_{z, 1 n}={\cal I}_x+i\,{\cal I}_y=\frac{ne\hbar}{2\pi\, m}\,\,\sqrt{\frac{m\,\omega}{ \hbar}}\,\,\,\frac{\Gamma(n+\frac{1}{2})}{\Gamma(n+1) }\,.
 \end{equation}
 Hence, no current flow along the y-direction. This situation corresponds to the case of Hall effect \cite{1}. The current is confined to flow  along the negative x-direction for a positive charge particle.  This current can be particularly useful for the study of  Hall current.
 It is interesting that the current along the x-axis is quantized and along the y-axis vanishes. Consequently, no current arises in the ground-state.
However, unlike the real harmonic oscillator where the current density vanishes, the current density for a complex harmonic oscillator is generally non-zero. If the current in eq.(19) arises from quasielectrons, then the current in eq.(21) should arise from quasiholes. It is apparent that each of the two states has two currents (current and its complex conjugated partner), and hence show two conductivities.
Thus, eqs.(6)- (9) will yield
 \begin{equation}\label{5}
{\cal I}_{z, 2 n}=e^*_{2 n}\,v_z\, \Longrightarrow {\cal I}_{2 n}=e^*_{2 n}\,\frac{<p_z>_{1 n}}{m}\,, e^*_{2 n}=\frac{n}{2n-1}\,e\,.
 \end{equation}
 and
\begin{equation}\label{5}
{\cal I}_{z, 1 n}=e^*_{1 n}\,v_z\, \Longrightarrow {\cal I}_{1 n}=e^*_{1 n}\,\frac{<p_z>_{1 n}}{m}\,, e^*_{1 n}=\frac{n}{2n+1}\,e\,,
 \end{equation}
 This implies that the effective electron (oscillator) charge, $e$, in a  state $\psi_{2 n}$ will be $e^*_{2n}=\frac{n}{2n-1}\,e$\, and in the state $\psi_{1 n}$ will be $e^*_{1n}=\frac{n}{2n+1}\,e$\,. Surprisingly, these are the same  fractional charge observed in the fractional quantum Hall effect (FQHE) \cite{laugh}. The charges $e^*_{1 n}$ and $e^*_{2 n}$ represent the charges of the quasiparticle in the two states, $\psi_{2 n}$ and $\psi_{1 n}$\,, respectively.   Laughlin  proposed that the elementary excitations of the FQHE state carry fractional charge \cite{laugh}. These fractional charges are those associated with  particle excitations (quasiparticles). For $n = 1, 2, 3, 4 \cdots$\,, one has the two conjugated series: $\frac{1}{3}, \frac{2}{5}, \frac{3}{7}, \frac{4}{9}\cdots $ and  $1, \frac{2}{3}, \frac{3}{5}, \frac{4}{7}, \cdots $\,, respectively. It is interesting to notice that the overlap between any two adjacent states, one with $\psi_{1 n}$ and other with $\psi_{2, n+1}$, produces an electron \cite{arbab}.
 An oscillation in the conductivity of a material that occurs at low temperatures in the presence of very intense magnetic fields is known as Shubnikov de Haas effect (ShdH) which is a macroscopic manifestation of the inherent quantum mechanical nature of matter \cite{shubni}.
 Thus, the free electrons in the conduction band of a metal, or narrow band gap semiconductor at very low temperatures and high magnetic fields,
behave like simple harmonic oscillators.
\section{The Hall conductivity, current, mobility and voltage}
 The Hall resistivity of $N$ electrons arising from the state $\psi_{2 n}$, and  $\psi_{1 n}$,  $R_H=\frac{V_y}{I_x}$\,, can be written as
 \begin{equation}\label{5}
R_{z2, H}=\frac{E_yW}{NI_{x, 2n}}=\frac{v_xBLW}{NI_{x, 2n}L}=\frac{v_zBA}{N{\cal I}_{z, 2n}}=\frac{v_z\phi_B}{N{\cal I}_{z, 2n}}\,,
 \end{equation}
 and
$$
R_{z1, H}=\frac{E_yW}{NI_{x, 1n}}=\frac{v_xBLW}{NI_{x, 1n}L}=\frac{v_zBA}{N{\cal I}_{z, 1n}}=\frac{v_z\phi_B}{ {N\cal I}_{z, 1n}}\,,
$$
 where $A=LW$ is the area of the sample, ${\cal I}_{z, 2n}=I_{z, 2n}L$ , and $\phi_B=BA$ is the magnetic flux.

The electronic  current for a system of $N$ electrons in the state $\psi_{1 n}$, as evident from eq.(20),  will be
\begin{equation}\label{5}
I_{z, 1n}=\frac{n_s\,e\hbar\,W}{2\pi\, m}\,\,\sqrt{\frac{m\,\omega}{ \hbar}}\,\,\,\frac{\Gamma(n+\frac{1}{2})}{\Gamma(n) }\,\,,
 \end{equation}
  where $n_s$ is the two dimensional electronic density. Notice that the current in the state $\psi_{1n}$ is equal to $I_{n1, z}=-I_{n2, z}$\,. Equation (26)
 can be written as
 \begin{equation}\label{5}
I_{z, 1 n}=I_0\,W\,\,\sqrt{\frac{m\,\omega}{ \hbar}}\,\,\,\frac{\Gamma(n+\frac{1}{2})}{\Gamma(n) }\,\,,\qquad I_0=\frac{e\hbar\, n_s}{2\pi\, m}\,\,.
 \end{equation}
The current $I_0$ is the characteristic (minimum) current resulted when the width of the sample is equal to the characteristic (magnetic) radius, $\ell_B=\sqrt{\frac{\hbar}{eB}}$\,,
when a magnetic field is applied. For a typical sample $n_s\sim 10^{15}m^{-2}$\,, we find that $I_0\sim 3\,nA$\,. For $B\sim 1\, T$\,, and $W\sim 1\, nm$\,, we see that $I_z\sim 1n\,A$\,, and for $W\sim 1\mu m$, one finds $I_z\sim 1\mu A$\,.

In two dimensions, the transverse resistance and transverse ($R_{xy}$) resistivity ($\rho_{xy})$, longitudinal resistance (magnetoresistance) ($R_{xx})$,
longitudinal resistivity ($\rho_{xx}$) are related by
 \begin{equation}\label{5}
R_{xy}=\rho_{xy}\,,\qquad \rho_{xx}=\frac{W}{L}\,R_{xx}\,.
 \end{equation}
 Applying  eqs.(23) and (24) in (25), the Hall  conductivity, $\sigma_{xy}=1/\rho_{xy}=1/R_{z, H}$, in the two states, $\psi_{2 n}$ and $\psi_{1 n}$, are respectively,
 \begin{equation}\label{5}
\sigma_{xy, 2n}=\frac{{N\cal  I}_{z, 2n}}{v_{z, 2n}\phi_B}=\left(\frac{n}{2n-1}\right)\,\frac{e^2}{h}\,,
 \end{equation}
 and the complex conjugate
 \begin{equation}\label{5}
\sigma_{xy, 1n}=\frac{N{\cal  I}_{z, 1 n}}{v_{z, 1n}\phi_B}=\left(\frac{n}{2n+1}\right)\,\frac{e^2}{h}\,,
 \end{equation}
 where $\phi_B=Ne/h$ is the flux quantum. One can accordingly define the  conductivity $\sigma_z$ as
 \begin{equation}\label{5}
\sigma_z=\sigma_{xx}+i\,\sigma_{xy}\,.
 \end{equation}
Similarly, the  resistance can be expressed as
\begin{equation}\label{5}
R_z=R_{xx}+i\,R_{xy}\,.
 \end{equation}
This implies that the resistivity will be
 \begin{equation}\label{5}
\rho_z=\rho_{xx}+i\rho_{xy}=\frac{1}{\sigma_z}=\frac{\sigma_{xx}}{\sigma^2_{xx}+\sigma^2_{xy}}-i\frac{\sigma_{xy}}{\sigma^2_{xx}+\sigma^2_{xy}}\,.
 \end{equation}
This yields the two relations
\begin{equation}\label{5}
\rho_{xx}=\frac{ \sigma_{xx} }{ \sigma^2_{xx}+\sigma^2_{xy} }\,,\qquad \rho_{xy}=-\frac{ \sigma_{xy} }{ \sigma^2_{xx}+\sigma^2_{xy} }\,.
 \end{equation}
 Hence, when $\rho_{xx}$ vanishes, $\sigma_{xx}$ will vanish too, and $\rho_{xy}=-\sigma^{-1}_{xy}$.

 The combined Hall resistance in the two states, $\psi_{1 n}(+)$ and $\psi_{2 n}(-)$, is given by
  \begin{equation}\label{5}
R_{z, H}=\left(\frac{2n\pm1}{n}\right)\,\frac{h}{e^2}\,,
 \end{equation}
 The longitudinal resistance (magnetoresistance) resulting from the state $\psi_{1 n}$ is given by
  \begin{equation}\label{5}
R_{xx}=\frac{2\pi\, mV_x}{n_s\,e\hbar\,W}\,\,\sqrt{\frac{ \hbar}{m\,\omega}}\,\,\,\frac{\Gamma(n) }{\Gamma(n+\frac{1}{2})}\,.
 \end{equation}
 From eqs.(28) and (36), the longitudinal resistivity reads
$$
\rho_{xx}=\frac{2\pi\, mV_x}{n_s\,e\hbar\,L}\,\,\sqrt{\frac{ \hbar}{m\,\omega}}\,\,\,\frac{\Gamma(n) }{\Gamma(n+\frac{1}{2})}\,.
$$
Using eq.(27), eq.(36) can be written as
  \begin{equation}\label{5}
R_{xx}=\frac{1}{I_0}\frac{V_x}{W}\,\,\sqrt{\frac{ \hbar}{m\,\omega}}\,\,\,\frac{\Gamma(n) }{\Gamma(n+\frac{1}{2})}\,\,.
 \end{equation}
Therefore, when $V_x\sim 1\,mV$\,, $B\sim 1\, T$, and $W\sim 1nm$\,, a resistance of $R_{xx}\sim 10^6\Omega$ will be produced.
 However, for $V_x\sim 1\,\mu\,V$, one has $R_{xx}\sim 1\, {\rm k}\Omega$\,.

 The corresponding Hall mobility is given by
 \begin{equation}\label{5}
\mu_H=\frac{2n_s\hbar\, L}{mV_x}\,\,\sqrt{\frac{\hbar}{m\,\omega}}\,\,\,\frac{\Gamma(n+\frac{3}{2})}{\Gamma(n+1) }\,.
 \end{equation}
The above two equations can be combined to give the Hall mobility as
 \begin{equation}\label{5}
 \rho_{xx}\,\mu_H=\frac{R_H}{B}\,.
 \end{equation}
One can now define the Hall voltage resulting from the state $\psi_{1 n}$\,, i.e., $V_H=I_zR_H$\,, as
   \begin{equation}\label{5}
V_H=\frac{2n_s\hbar^2W}{m\,e}\sqrt{\frac{m\omega}{\hbar}}\,\,\,\, \frac{\Gamma(n+\frac{3}{2})}{\Gamma(n+1)}\,.
 \end{equation}
Using eq.(40), it is interesting to express the Hall's electric field as
\begin{equation}
E_H=\frac{2n_s\hbar^2}{m\,e}\sqrt{\frac{m\omega}{\hbar}}\,\,\,\, \frac{\Gamma(n+\frac{3}{2})}{\Gamma(n+1)}\,.
 \end{equation}
Using eq.(27) in eq.(41), one finds
\begin{equation}
E_H=\frac{1}{\tau_s}\,\sqrt{\frac{\hbar\, B}{e}}\,\frac{\Gamma(n+\frac{3}{2})}{\Gamma(n+1)} \,,\qquad \tau_s=\frac{m}{\hbar \, n_s}=\frac{e}{2\pi I_0}\,.
 \end{equation}
In a real two - dimensional system, like 2-dimensional electron gas (2DEG), electrons are trapped in a thin layer ($\sim 10\, nm$)  made at the interface between a semiconductor and an insulator, or between semiconductors. A typical device is the GaAs/GaAlAs heterostructure \cite{laugh}. The transition occurs when the Landau level crosses the Fermi level. In this instant, one has
 \begin{equation}
 \epsilon_F=(n+1)\hbar\omega_c\,,\qquad \omega_c=eB/m\,,
 \end{equation}
where in two-dimensions, the Fermi energy ($\epsilon_F$) is given by (ignoring electron' spin)\footnote{In case of considering the electron's spin, $n_s\rightarrow n_s/2$.}
 \begin{equation}
 \epsilon_F=\frac{2\pi\hbar^2n_s}{m}\,.
 \end{equation}
 Hence, under the application of the magnetic field, we obtain
 \begin{equation}
 n=\frac{n_sh}{eB}-1\,,
 \end{equation}
 where $n$ is the principal quantum  umber of the oscillator.  Equation (45) shows clearly that not all magnetic fields can give rise to transition; only those which are compatible with eq.(45) are allowed. In general, $eB/h\le n_s$\,. One can now define the magnetic number density of electrons as, $n_B=eB/h$. Thus, the 2DEG number density must exceed the magnetic number density, i.e.,  $n_s>n_B$. Hence, a magnetic field of $10\,T$ requires $n_s\ge 2.4\times 10^{15}m^{-2}\,$ before the quantization is observed. Notice that $\frac{n_sh}{eB}$ must be an integer, i.e., $\frac{n_sh}{eB}=i$, where $i$ is an integer. This is the condition (filling factor) that Klitzing has found for the quantized two-dimensional Hall's resistance \cite{kilit}. Hence, eq.(45) now reads that $i=n+1$. Therefore, the filling factor $i$ is a signature of the principal quantum number of the harmonic oscillator.

The two-dimensional classical Hall's resistance is given by $R_H=\frac{B}{n_se}$, which yields the quantum Hall's resistance $R_H=\frac{h}{ie^2}$. Thus, the Hall's resistance in terms of the principal quantum number, $n$, is
 $$
R_H=\frac{1}{n+1}\,\frac{h}{e^2}\,,\qquad\qquad n=0, 1, 2, 3, \cdots.
 $$
Thus, no quantization can take place unless the magnetic field $B=\frac{n_sh}{e\,i}$ is reached. This quantization is exhibited in the longitudinal and transverse resistances pattern observed in the FQHE \cite{tsui}. However, other values of $B$ lead to the plateau pattern. In these situations the Fermi level lies between two Landau levels.
It seems that the Hall's resistance is quantized whenever the Landau level crosses the Fermi level.  It is apparent from eq.(45) that as the magnetic field starts to increase, the Landau energy levels sweep from infinity to the ground state. This sweep induces oscillations in the longitudinal resistance and resistivity. To this aim, $R_{xx}$, $\mu_H$ and $V_H$ can be evaluated for a 2DEG in conjunction with the condition in eq.(45).
Notice that applying eq.(45) in eq.(35) implies that $R_H\rightarrow 2h/e^2$\, as $B\rightarrow 0$\,, while $R_H\rightarrow h/e^2$\, as $B\rightarrow \infty$\,. Therefore, the Klitzing resistance is reached (in the FQHE) when a very high magnetic field is applied. This shows that the higher states (Landau levels)  cross  the Fermi level first, and then follow the remaining lower states.

The width of the plateau can be obtained from eq.(45), \emph{viz.},

\begin{equation}
\Delta B=\frac{1}{n(n+1)}\,\frac{n_sh}{e}\,\,\qquad n=1, 2, 3, \cdots\,.
\end{equation}
 Hence, the width of the plateau between the ground state ($n=0$) and the first excited state ($n=1$) is $\frac{n_sh}{2e}$\,, which is the maximum width. Apparently, the factor $\frac{n_sh}{e}$ is a measure of the plateau width. It is apparent that the width of the plateau is a decreasing function of the principal quantum number of the oscillator, $n$\,.

 Now the Hall's derivative with respect to the magnetic field can be obtained from eq.(35) employing eq.(45), which is
$$
\frac{dR_H}{dB}=\pm\frac{1}{e\,n_s}\,\frac{1}{(1-\frac{eB}{n_sh})^2}\,,
$$
where $\pm$ stands for $\psi_{1n}$ and $\psi_{2n}$\,, respectively.
For a small magnetic field, the above equation  reduces to the classical case, \emph{i.e.}, $R_H\rightarrow B/(n_se)$. However, for large magnetic field, $\frac{dR_H}{dB}\rightarrow \pm(\frac{n_sh}{eB})^2\,$, \emph{i.e.}, as $B\rightarrow\infty$, $\frac{dR_H}{dB}\rightarrow 0$, so that $R_H$ will flatten. This would explain why we observe a plateau in the Hall's resistance.
For small magnetic field, the above equation can be expressed as
$$
\frac{dR_H}{dB}=\frac{1}{e\,n_s}\exp(-\frac{2eB}{n_sh})\,.
$$
To observe the temperature effect, we write the above equation as
$$
\frac{dR_H}{dB}=\frac{1}{e\,n_s}\exp(-\frac{\Delta E}{k_BT})\,,\, \Delta E=\hbar\,\omega_c\,,\, T_H=\frac{\pi\hbar^2n_s}{k_Bm}\,,
$$
where $T_H$ is the transition temperature (Hall's temperature) from a given state to another, which is about few Kelvins for 2DEG. This shows clearly that the FQHE is a low temperature phenomenon. We can now relate this temperature to the Heisenberg's uncertainty relation, $\Delta E\Delta t=h/2$, where $\Delta E=k_BT_H$ and $\Delta t=\tau_s$.

 \section{Symmetries of the FQHE}
We remark that there are two states in any energy level, these two states have different charges and hence different conductivities. I prefer to call the charges of these states [eqs.(23) \& (24)] the conjugated charges, and the corresponding conductivities [eqs.(29) \& (30)], the conjugated conductivities. Interestingly, these are the two observed sets of conductivity in the FQHE \cite{tsui}. The number of states within each Landau level (when is full) and the Hall conductivity are  given by $N_L=n_B=\frac{eB}{h}\,$ and $\sigma_{xy}=-\frac{Ne}{B}\,$, respectively \cite{landau}. Hence, combining eqs.(23) and (24) with (29) and (30) yields
  \begin{equation}\label{5}
\frac{N}{N_L}=\frac{n}{2n\pm 1}\,.
 \end{equation}
 The ratio in eq.(47) defines the filling factor in the FQHE.
 Interestingly, if we now allow $n\rightarrow -n$, then $e_{2n}^*\leftrightarrow e^*_{1n}$ and $\sigma_{xy, 2n}\leftrightarrow \sigma_{xy, 1n}$. This means the two states $\psi_{1 n}$ and $\psi_{2 n}$, are (conjugate) compliment to each other.  Thus, we can associate with this  transformation, of one state into another, some physical symmetry (complementarity).  Recall that the corresponding orbital angular momentum of the states, $\psi_{2 n}$ and $\psi_{1 n}$, are respectively, $n\hbar$ and $-n\hbar$. However, the two states have the same energy of $(n+1)\,\hbar\,\omega$.  They thus represent a bound (condense) particle. Therefore, in all states one has such a condensed state. Moreover, as $n\rightarrow \infty$, $e^*_{1n, 2n}\rightarrow \frac{1}{2}\, e$\,, in both states. Thus, the entire two dimensional oscillators (gas) is in a state of Bose-Einstein condensation at low temperature.

 One may consider some dual transformation where, $n\rightarrow 1/n$. Under this transformation,
  \begin{equation}
 E_n\longrightarrow \frac{E_n}{n}\,,\qquad \qquad L\longrightarrow \frac{L}{n}\,\,.
 \end{equation}
  Thus, the dual transformation is equivalent to a scaling of energy and angular momentum.
 The conductivity in the two states, $\psi_{1 n}$ and $\psi_{2 n}$, under the dual transformation will be
  \begin{equation}\label{5}
\sigma_{xy, 1n}=\left(\frac{1}{2+n}\right)\,\frac{e^2}{h}\,,\qquad  \sigma_{xy, 2n}=-\left(\frac{1}{2-n}\right)\,\frac{e^2}{h}\,,
 \end{equation}
 respectively. Hence, the range of conductivity is enlarged.

It seems that as the magnetic field increases and the temperature is lowered, electrons tend to form condensed states. The system will thus behave as a bosonic system. The  electron can be formed from the supperposition of two quasiparticles states; one orbiting clockwise in a given state ($n$) and the other orbiting counterclockwise in the  state next to it ($n+1$). Thus, the electron wavefunction can be written as a superposition of the two states: $\psi_{2 n}$ and $\psi_{1, n+1}$\,, respectively. In such a case the electron will have a charge of $e$\,. In the FQHE the ratios $\frac{n}{2n\pm1}$ correspond to the filling factor \cite{laugh}.

 Let us now consider the effect of the transformation $z\rightarrow e^{i\theta}z$ (rotation of the complex coordinates) on the wavefunctions, $\psi_{2 n}$ and $\psi_{1 n}$. Note that the Hamiltonian and the orbital angular momentum are invariant under this transformation. However,
\begin{equation}
\psi_{2 n}\rightarrow e^{in\theta}\psi_{2 n}\,\,,\qquad \qquad \qquad \psi_{2 1}\rightarrow e^{-in\theta}\psi_{2 1}\,.
 \end{equation}
 It seems that when the coordinates are rotated in certain direction, the state $\psi_{2 n}$ is rotated in the same direction whereas  $\psi_{1 n}$ is rotated in the opposite direction.
 Thus, the wavefunctions preserve their states when rotated by the angles
 \begin{equation}
\theta_1=\frac{2\pi}{n}\, N_w\,,\qquad\qquad  N_w=0, \pm1, \pm2, \pm, \cdots\,,
 \end{equation}
 where $N_w$ is the number of windings.
However, if we want to make the effect of rotation equal to reflection (where particles are exchanged), then the corresponding rotation angles are
  \begin{equation}
\theta_2=\frac{\pi }{n}\,(2N_w+1)\,,\qquad\qquad  N_w=0, \pm1, \pm2, \pm, \cdots\,.
 \end{equation}
 We see that the ratio between the two above angles is $\frac{\theta_1}{\theta_2}=\frac{N_w}{2N_w+1}$. Since $N_w$ takes integer numbers, we may associate this ratio to the ratio of charges of the two states. This may explain the observed fractional parameter ($\nu$) in the FQHE \cite{laugh}.

Because of Pauli exclusion principle, fermions wavefunctions should be antisymmetric under the exchange of coordinates (particles).
We remark here that, under the above transformations, the energy of the states doesn't change.

In Aharonov-Bohm effect  the complex phase of a charged ($q_0$) particle's wavefunction  couples to the electromagnetic potential ($\vec{A}$)\cite{bohm}. Consequently, the particle's wavefunction experiences a phase shift as a result of the enclosed magnetic field ($\vec{B}$) when passing around a long solenoid. Thus, an observable shift of the interference fringes  will then occurs. The same phase effect is responsible for the quantized-flux requirement in superconducting loops. Thus the phase difference  around a closed loop must be an integer multiple of $2\pi$ (with the charge for the electron Cooper pairs), and thus the flux $\phi_B$ must be a multiple of $h/2e$ \cite{london}.  The number of times the closed path  encircles the solenoid is defined by a winding number which is a topological property of the space in which the charged particle moves. This amounts to say that the Aharonov - Bohm effect is a quantum mechanical topological effect.  Accordingly, one has
\begin{equation}
\psi\rightarrow e^{ \frac{iq_0}{\hbar}\int \vec{A}\cdot d\vec{\ell}}\,\,\,\psi\,\,,\qquad \qquad \vec{B}=\vec{\nabla}\times\vec{A}\,.
\end{equation}
Hence, the phase shift will be
 \begin{equation}
\Delta\varphi=\frac{q_0\phi_B}{\hbar}\,,\qquad \qquad \phi_B=\int \vec{A}\cdot d\vec{\ell}\,.
\end{equation}
This phase shift has been observed experimentally \cite{osaka}. The two wavefunctions in eq.(51) interfere when one particle is rotating clockwise and the other counterclockwise. This may mimic the effect of the Aharonov - Bohm effect.

If  we now treat the transformation of the wavefunctions in eq.(52) as the on resulting from the  rotation of the complex coordinates, i.e., as in eq.(23), then eq.(54) would imply
 \begin{equation}
\theta=\frac{q_0\phi_B}{n\hbar}\,.
\end{equation}
For a unit flux quantum, i.e., $\phi_B=h/e$, one has $\theta=\frac{2\pi}{2n\pm 1}$. Hence, rotating the complex coordinates, which rotates the wavefunctions, is equivalent to introducing a gauge transformation. Hence, this usher to  some kind of interaction between the oscillators and a photonic field described by $\vec{A}$.
For constructive interference (of the two states: $\psi_{1 n}$ and $\psi_{2 n}$), we have $\theta=2\pi\, N_w$ and hence
\begin{equation}
\phi_B=N_w\,\frac{h}{q_0}\,\,,\qquad \phi_B=N_w\,\frac{2n\pm 1}{n}\, \frac{h}{e}\,, \qquad N_w=1, 2, 3, \cdots\,.
\end{equation}
This intriguing result shows that the magnetic flux is quantized. Furthermore, a flux quanta is associated with each charge (particle). It seems that the FQHE  is not a mere quantum mechanical effect but rather a quantum electrodynamics one.  Moreover, note that when two particle are (rotated) exchanged their wavefunctions acquire the  Aharonov-Bohm phase \cite{bohm}.

\section{Acknowledgments}
I thank Drs. R. Shah, H. M. Widatallah and M. A. Khalafalla for fruitful  discussions  and critical comments.  I would like to express my thanks to Sultan Qaboos University (Oman) for visiting grant where  this work has been carried out.

\end{document}